\documentclass[fleqn,10pt]{wlscirep}
\usepackage[utf8]{inputenc}
\usepackage[T1]{fontenc}
\usepackage{float} 
\title{Inertial Self-Caging: Dynamics of Macroscopic Swimmers at Moderate Reynolds Number Sustaining Chemical Wake Resonance}

\author[1]{Alessandro Foradori}
\author[1,*]{Paolo Bettotti}
\affil[1]{Department of Physics, Nanoscience Laboratory, University of Trento, via Sommarive 14, 38123 Povo(TN), Italy}

\affil[*]{paolo.bettotti@unitn.it}


\begin{abstract}
Self-propelled phoretic swimmers are generally studied in the laminar flow regime, where their low speed renders inertial effects negligible and trajectories highly predictable. This research tackles the challenge of propulsion in the inertial regime, at moderate Reynolds numbers ($Re \sim 100-200$), where fluid dynamics becomes non-linear. By using a chemically driven macroscopic hydrogel this work demonstrates, through experiments and modeling, the existence of stable resonant states under confined geometry: as the swimmer circles, it interacts with its own lasting chemical wake. This chemical self-feedback creates a complex, stable motion characterized by both universal exponential speed decay with superimposed significant periodic speed oscillations. Furthermore, a critical threshold speed is identified, where the system abruptly transitions from the resonant oscillatory regime to a stochastic stop \& go behavior. These findings provide a fundamental understanding of how chemical fields, hydrodynamic inertia, and confinement couple to determine the motion properties of high-speed active matter.
\end{abstract}
\begin{document}

\flushbottom
\maketitle

\thispagestyle{empty}

\section*{Introduction}
The physics of active matter has emerged as a cornerstone of non-equilibrium science, offering insights into biological locomotion \cite{Liu2021Viscoelastic, Ross2019Controlling, Prakash2023Dynamics}, swarming behaviors \cite{Casiulis2024Geometric, flexicles}, and the design of autonomous microrobots \cite{Palagi2018113, Shahsavana2024Bioinspired, Cerona2024Programmable, haidong2025}. Historically, the theoretical foundation for the analysis of the dynamics of synthetic swimmer motion rests firmly in the realm of low-Reynolds number hydrodynamics ($Re \equiv uL/\nu \ll 1$ where $Re$ is the Reynolds number, $u$ is the speed of the swimmer, $L$ is the characteristic swimmer length and $\nu$ is the kinematic viscosity). In this regime, viscous forces dominate, and the time-reversible nature of Stokes flow imposes stringent limitations on achieving propulsion, often necessitating complex nonreciprocal actuation to break time-reversal symmetry (the so called Scallop theorem) \cite{Purcell1977Life}. The scaling up of active particles to the macroscopic scale, or their engineering for high-speed operation, fundamentally challenges this paradigm by transitioning the dynamics into an intermediate regime where the inertial term of the full Navier-Stokes (NS) equations ($\rho(\mathbf{u} \cdot \nabla) \mathbf{u}$) is no longer negligible, and it introduces time-irreversibility and complex flow structures (wakes). The intermediate Reynolds regime ($Re \sim 100-500$) is of interest in many natural systems (i.e. to describe the motion of small crustaceans, tiny insects and blood flow in small arteries) as well as for  technological applications (i.e. microfluidic mixers, ink jet printers). While the theoretical study of non-linear inertial hydrodynamics is a massive challenge in itself, the complexity is compounded when the motion is coupled to a dynamic chemical field, as described in the present paper where Marangoni Effect (ME) is exploited to generate thrust via surface tension gradients. In fact, ME-driven swimmers are uniquely capable of generating the forces required to achieve sustained, macro-scale motion at high $Re$, distinguishing them from other chemically powered systems that often remain in the laminar domain ($Re \ll 1$). Accurate modeling of this moderate-$Re$ ME propulsion requires solving the nonlinearly coupled NS equations with the Advection-Diffusion equation, which governs the transport of the released chemical fuel; such coupled system is notoriously difficult and requires substantial computational approximations, sometimes arriving at absurd results\cite{Boniface2020}.
The complex dynamics of the phoretic swimmer in the intermediate $Re$ regime has been investigated by only a few groups and it is still an active topic: generally, the experiments are conducted using a material that either dissolves in time (the most striking example being the camphor boats) or using partially miscible liquids that do not saturate the liquid surface. Notably, all these experiments 
tend to reduce the degrees of freedom allowed for the motion of the swimmer by confining it into special containers or forcing it to move along a circular trajectory using mechanical constraints\cite{Tanaka2015, D2CP05707G, cam_nakata2023}.
Letting the swimmer move freely, albeit within a spatially limited domain, permits the investigation of all types of motion, some of which are not accessible by adding external constraints.
While oscillatory fluctuations in swimmer speed have been previously noted by others\cite{PhysRevResearch.2.043299,AKELLA20181176}, their origin has never been discussed in details. Our findings clearly establish, both experimentally and theoretically, that the oscillations are the result of a dynamic self-interaction: the swimmer moves through, and reacts with, the localized, incompletely mixed chemical cloud of surfactant it has just released into the pool solvent.
The interest in studying these resonances is twofold: 1) their existence indicates a peculiar resonant condition that severely biases the swimmer's motion and 2) since they appear when the swimmer's thrust is maximal (ie. in the initial part of the motion), they have a critical impact on the resulting dynamics and, thus, on the processes the swimmer is programmed to perform.
This level of control and reproducibility is the key advancement, allowing us to move beyond complex, data-driven approaches and instead build physically insightful, minimal models. While capturing the intricate details of the surrounding fluid flow would necessitate resolving the full 3D Navier-Stokes equations, this system operates in an intermediate regime where chaos has not yet fully developed. This crucial observation allows us to posit that the dominant inertial effects can be captured using a simplified, yet powerful, set of coupled equations of motion.\\

\section*{Results}
The swimmer is a cylinder ($10\ mm$ in diameter and $3 mm$ thick) made of nanocellulose hydrogel immersed in an alcoholic solution and stained with a dye: ethanol acts as surfactant, while the dye is required to track swimmer movement. The swimmer is released into a circular pool filled with solutions of water/glycerol with different concentration ratios, allowing for the variation of the viscosity of the medium. To accurately and reproducibly control the swimmer's dynamics, a simple system is implemented that balances buoyancy and surface tension forces in order to exploit the \emph{Cheerios effect}\cite{Vella2005Cheerios}. Specifically, the pool is filled beyond the rim with water/glycerol solutions, forming a convex meniscus. Conversely, since the swimmer is hydrophilic, the local meniscus surrounding it is concave and it experiences a repulsive force whenever its straight trajectory tends to bring it close to the liquid/air convex meniscus, which effectively constrains it to stay away from the edge. The motion of the swimmer is recorded on a digital camera and its trajectory is reconstructed by analyzing the video. The typical experiment shows two regimes of motion: an initial exponential decrease of the swimmer speed overlapped to a significant sinusoidal modulation of the speed itself (called it \emph{"oscillatory"} part). This regime is interrupted by an abrupt stop of the motion and is followed by an intermittent one, which is named \emph{"stop \& go"} regime (such a regime was already noted in \cite{PhysRevE.105.014216}).

\begin{figure}
    \centering
    \includegraphics[width=0.8\columnwidth]{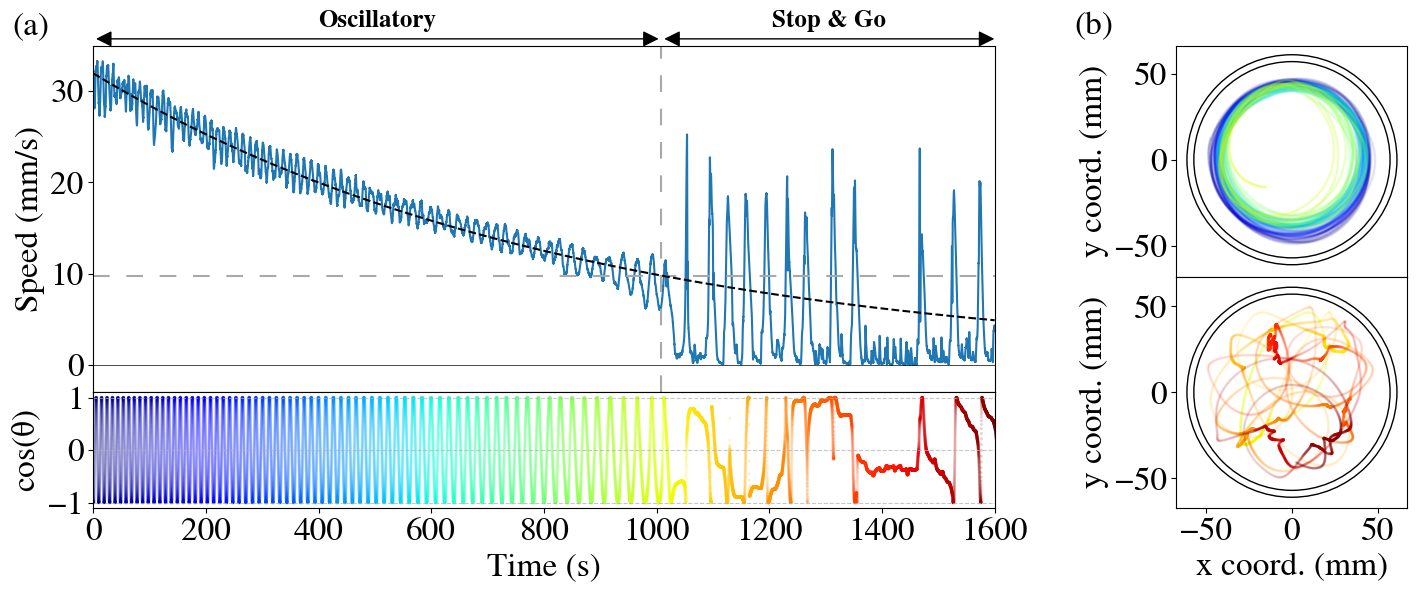}
    \caption{Typical dynamics of a swimmer motion. (a-top) Profile of the speed of the swimmer. Initially the speed follows an exponential oscillating decay with superimposed significant oscillations. After a certain (and variable) time the swimmer comes to an abrupt stop and then it enters into a stochastic, stop \& go, regime. (a-bottom) The rainbow line reports the cosine of the angle of the swimmer position with respect to $\hat{u_{x}}$ centered at the pool center. During the initial phase of the motion the swimmer performs closed trajectory following the edge of the container and they are characterized by a chirped period of the oscillations. (b) Typical trajectory of a swimmer: (Top) During the exponential decaying phase the inertial movement propels the swimmer along the pool wall edges. (Bottom) When the speed reaches a critical threshold value the motion stops abruptly and a wandering trajectory develops. The rainbow-coded color corresponds to the one of time traces in panels (a) and (b).}
    \label{fig:speed_vs_time}
\end{figure}
The result of a typical experiment is shown in Fig. \ref{fig:speed_vs_time}(a) the speed of the swimmer ($\mathbf{v}(t)$) is reported in blue. For this specific experiment, up to about $1000\ s$ the speed follows an exponential decay (fitted by the black dashed line) with superimposed large speed modulations. The bottom panel reports the cosine of the angle ($\theta (t)$) that the actual position of the swimmer ($\mathbf{r}(t)$) makes with an horizontal, right oriented reference x-axis ($\hat{u_{x}})$ (the origin of the coordinate system is at the pool's center). The vertical gray dashed line indicates the breakpoint that divides the oscillatory from the stop \& go regime. The horizontal dashed gray line indicates the $v(t)$ at breakpoint. A strong correlation between the speed oscillations and the cosine of the angle is clearly evident in the oscillatory part. For times beyond the breakpoint, both trajectory and speed assume chaotic behavior. Fig. \ref{fig:speed_vs_time}(b) reports the spatial trajectory of the swimmer and the concentric black circles indicate the thickness of the pool wall. The spatial trajectories are color-coded by the elapsed time, using the same spectral colormap applied to $cos (\theta (t))$.
Notably, the sudden stop of the swimmer after the oscillating behavior is unusual since an inertial object, moving into a viscous fluid, should decrease its speed asymptotically to zero unless a threshold due to nonlinear interactions develops. Thus, a precise determination of the breakpoint time is crucial to separate the two regimes and investigate the physical origin of the oscillatory one. Typical change point detection methods \cite{ruptures, stumpy} are not robust enough to locate it and two alternative methods were developed. They can be used interchangeably for its detection by exploiting the different symmetries of the trajectories in the two regimes, with the first part showing nearly circular trajectories and the latter one showing chaotic movements. They are described in detail in the ESI. Essentially, the curvature of each closed loop is analyzed and an empirical value of $10^{-2} mm^{-1}$ is used as an effective threshold to define the time at which the motion changes and it is valid for all fluid compositions investigated (different water/glycerol ratios): the first trajectory having local curvature above such threshold defines the start of stop \& go motion, dominated by chaotic accelerations.

\begin{figure}
    \centering
    \includegraphics[width=.8\linewidth]{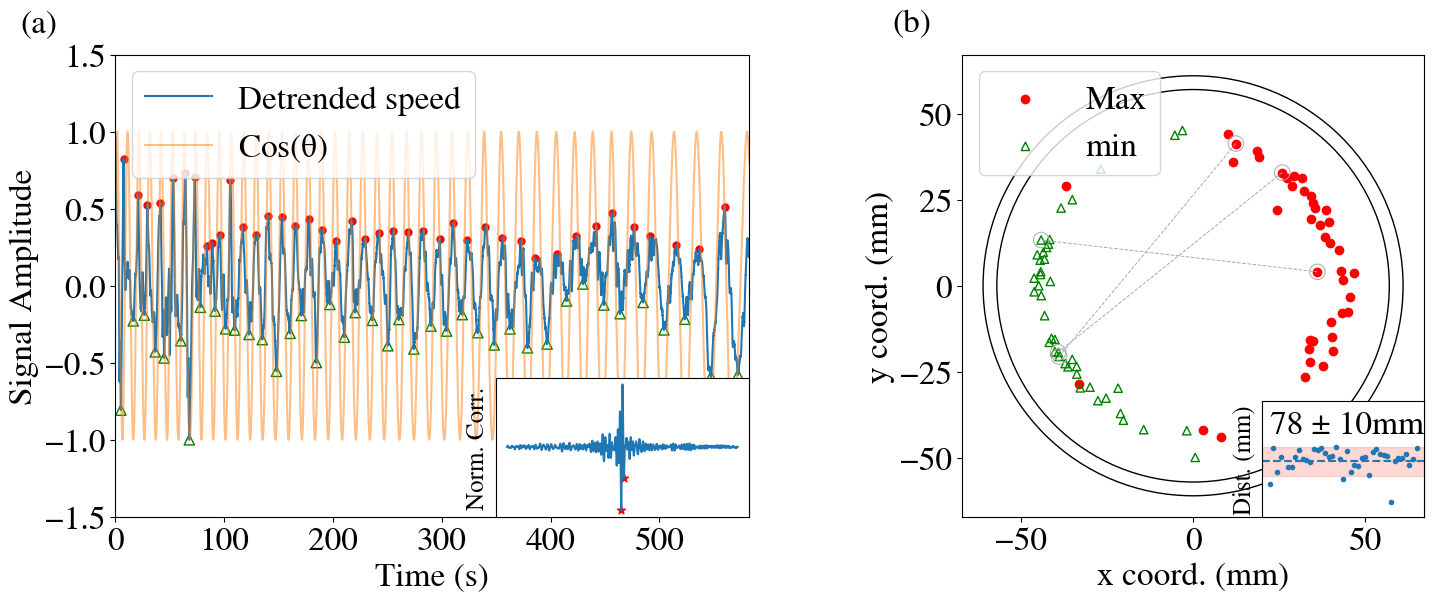}
    \caption{Analysis of both temporal and spatial cross correlation between the speed and position of the swimmer. (a) Time correlation of the oscillations of the speed (detrended by subtracting the exponential decay, blue line) and cosine of the angular position of the swimmer (orange line) are overlapped and show a clear correlation. The inset shows the cross correlation of these signals; the two small red dots mark the minima of the cross correlation that correspond to $\pm \pi$ shift from the average period of the oscillations. (b) Spatial correlation between max and min of the speed. Symbols refer to same quantities of panel (a). Generally max and min tends to space apart but there is a significant noise. The three gray lines map three adjacent couples of max and adjacent minima. The inset shows the dispersion of the separations between adjacent max and min: the horizontal line marks the average value ($80 \pm 10\ mm$) and the shaded region is $\pm 1 \sigma$.}
    \label{fig:time_space_corr}
\end{figure}

Fig.\ref{fig:time_space_corr}(a) overlaps the detrended speed (ie. subtracted by the exponential decay, blue line) and the $cos(\theta)$ values (orange line).
The resonant effect between the swimmer's speed and the cosine of the angle is clearly evident in this part of the trajectory together with a significant chirp of the frequency of the oscillations. This effect is due to the reduced flux of ethanol released by the swimmer as time passes by. Red dots and green empty circles mark the extrema used to calculate their normalized cross correlation, which is shown in the inset: the strong positive peak at lags near zero confirms the correlation and two negative peaks (labeled with red stars) correspond to an advancing (delaying) of $\pi \ rad$ and their separation compares very well with the average period of the oscillations. While the temporal correlation is evident, the spatial one shows a greater amount of stochasticity. Fig.\ref{fig:time_space_corr}(b) shows the swimmer positions at the max (red dots) and min (green empty triangles) of the speed oscillations. Although successive pairs of maximum and minimum speed values are clearly localized at diametrically opposed positions in the pool, there is significant uncertainty in predicting their exact locations. The inset shows the average distance between adjacent max-min speed values (i.e. those within the same circular loop): the average value compares well with the mean diameter of the loops and its Coefficient of Variation ($CV = \sigma(x) / \bar{x}$)) is generally $CV < 20\%$. On the other hand, the distance between adjacent speed max (or min) is random: although they share the same typical length-scale (with values roughly about few mm), their $CV \sim 100\% $ indicates a stochastic process. 

\begin{figure}
    \centering
    \includegraphics[width=0.8\linewidth]{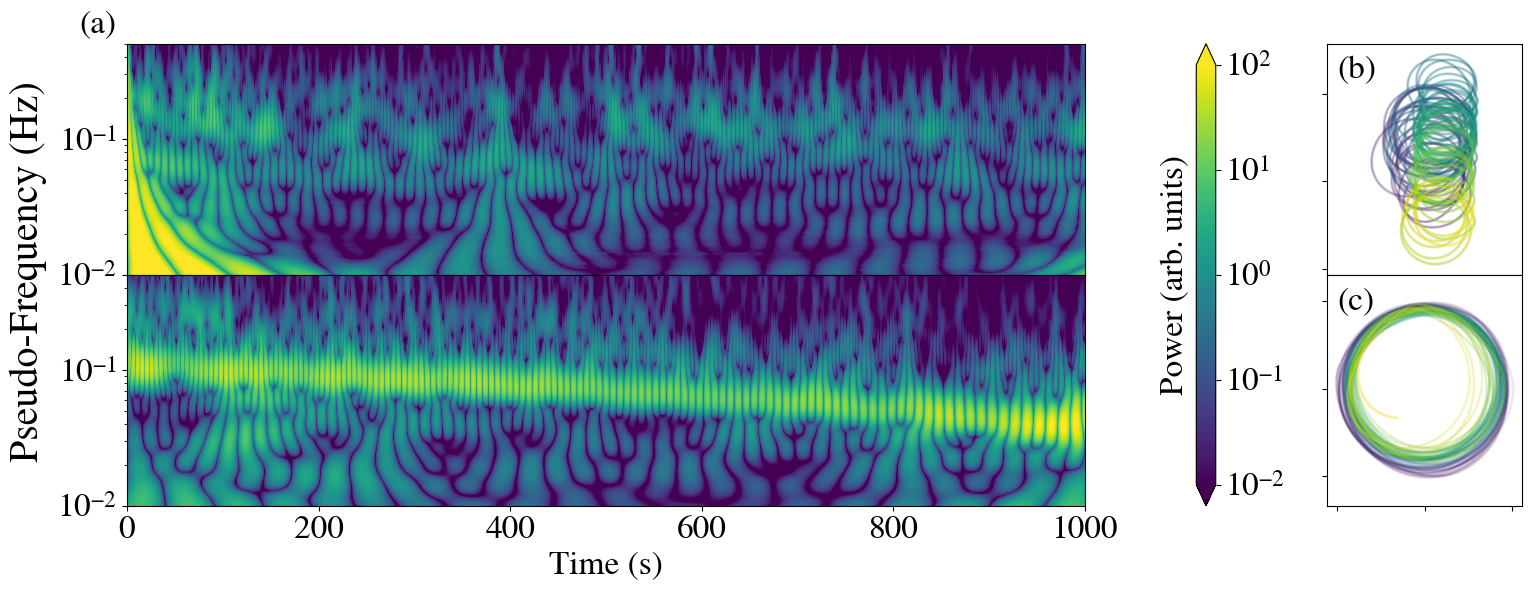}
    \caption{(a-top) Spectrogram of the speed oscillations for a swimmer moving in a large pool. The trajectory that creates the spectrogram is shonw in (b). (a-bottom) Spectrogram for a swimmer confined inside a circular pool. The trajectory for this particular experiment is shown in (c). No pool outline was added to either (b) or (c) panel since the large size of the square pool would prevent a proper visualization of the details of the trajectory. In the large square pool, the swimmer explores an area of about $50 \times 25\ cm$, while the round small pool is slightly larger than the average circular loop (see the top panel of Fig.\ref{fig:speed_vs_time}(b)).}
    \label{fig:wavelets}
\end{figure}

Such a resonant dynamics does not happen in large pool, where the swimmer is allowed to explore a larger space and it is not force to transit over its own chemical cloud. Experimental data on large pool and the comparison with the confined system is shown in Fig.\ref{fig:wavelets}.

Fig.\ref{fig:wavelets}(a) top panel shows the spectrogram of the speed oscillations for a swimmer released in a large pool (50 x 70 cm), the depth of the water is similar to that of the confined system. The bottom panel is the spectrogram of a swimmer in the circular pool. Fig.\ref{fig:wavelets}(b) shows the trajectory of the swimmer in the large pool, while Fig.\ref{fig:wavelets}(c) the trajectory in the small one. The large container prevents the possibility to obtain a convex meniscus thus, in this case, the swimmer tends to adhere to the pool wall because of the concave meniscus. This fact creates the right alignment of the closed trajectories that are attracted by the pool wall. Despite the lack of physical confinement, the swimmer propels, forming closely packed loops while slowly migrating back and forth around the region already explored. This fact might arise from a locally reduced liquid viscosity produced by the release of ethanol in these areas and it supports the idea of a local (and possibly temporal) non ideality of the environment in which the swimmer moves, which might be related to the sudden drop in the swimmer speed at threshold.

Concerning how the decay rate is affected by the solvent viscosity, the results show that water/glycerol ratios (W/G) larger than those investigated hamper the motion: the swimmer perform few circular loops before entering in an almost quiescent state; on the other hand, at low viscosity the probability to get a stable motion decreases and, most often, the trajectories evolve into either metastable or random paths (see a more in depth discussion in the ESI). The results indicate that the probability of having a stable motion of the swimmer (ciruclar loops around the pool edge lasting for hundreds of seconds) is inversely related to the solvent viscosity: for W/G=3/1, 4/1 and 5/1, the probabilities are of 71\% (10/14), 50\% (6/12) and 24\% (4/17), respectively (the number in parenthesis denotes the number of successful experiments vs the total number performed for that specific W/G ratio).
\begin{figure}[!b]
    \centering
    \includegraphics[width=0.3\linewidth]{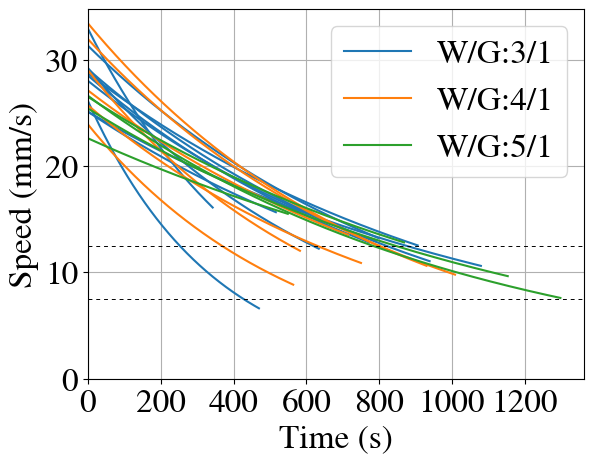}
    \caption{Interpolated decay of the speed for experiments in which oscillations last for at least few hundreds of seconds (some tens of loops). The lines color refer to specific W/G ratio as indicated in the legend. Generally, decays stop abruptly at speed threshold values of about $10 \pm 2.5\ mm/s$.}
    \label{fig:speed_decay}
\end{figure}
Fig.\ref{fig:speed_decay} resumes the decrease in speed with time during the oscillatory regime. The initial speed shows a variation of about 30\% that is the combined effect of the different viscosity of the solutions (between $1.75 \cdot 10^{-3} Pa \cdot s$ and $2.42 \cdot 10^{-3} Pa \cdot s$) and of the limited control over the release of the swimmer in the pool: in fact, the initial instants of the motion are highly unreproducible due to the uncertainty on how the surfactant diffuses around the swimmer before inertia prevails. The dispersion of initial speeds is of the same order as the variation in viscosity; this high variance obscures any potential trend between speed and solvent viscosity. Notably, despite the noisy behavior of the initial value of the speed, if the oscillatory regime lasts for at least a few hundreds of seconds (i.e. roughly 30 closed loops) the speed decreases smoothly down to a common threshold of about $10 \pm 2.5\ mm/s$ and then it stops abruptly when the stop \& go motion starts. The time at which the motion changes from oscillatory to stop \& go depends on the experiment and this variability reflects the stochasticity of the motion due to the complex interplay between diffusion and advection.

The invariance of this threshold speed across different W/G ratios suggests that it is independent of both the duration of motion and the cumulative amount of ethanol released (excluding a few of the shortest trials where structural defects likely induced pathological dynamics). This consistency implies that the threshold speed is governed by a nonlinear interaction between the swimmer's inertial effects and its immediate local environment. Indeed, the liquid interface is characterized by an excess concentration of ethanol \cite{Kim2017Solutal, Phan2021Surface, Doppelhammer2023Generation, Kirschner2021Molecular}, the mixing dynamics of which are fundamental to the ME. Furthermore, while a surfactant source in a pure water pool can induce a \emph{transparent zone} with Marangoni flow velocities an order of magnitude higher than those recorded here \cite{PRLMarangoni}, the swimmer's geometry poses important constraints to the dynamics. Specifically, as the swimmer is submerged to a depth of approximately 30\% of its diameter, the recorded velocities likely emerge from a delicate balance between the ME-driven \emph{pulling} at the liquid-air interface and the opposing viscous drag acting on the submerged body, potentially supplemented by diffusiophoretic contributions.

\begin{figure}
    \centering
    \includegraphics[width=0.8\linewidth]{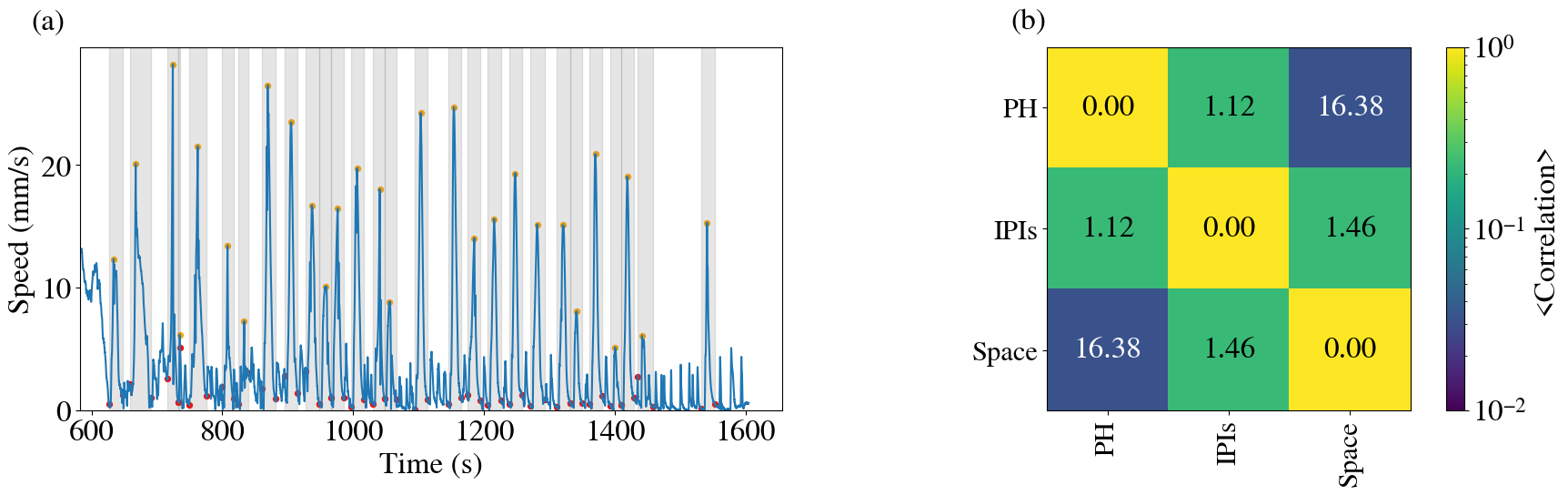}
    \caption{(a) Typical dynamic of the speed of the swimmer in the stop \& go regime. Orange dots indicate speed peaks and red dots the adjacent local minima. Gray area are the widths used to integrate the peak area. Note that the time scale for the stop \& go regime starts at breakpoint. (b) Mean lag-1 correlation for fundamental parameters of motion averaged over the entire set of experiments for W/G=4/1; PH: Peak Height, IPIs: Inter Peaks Intervals, Space: space traveled during the event (in mm). The colors of the map report the values of the correlations, while the numbers indicates the CVs: (IPIs, PH) and (Space, IPIs) are vaguely correlated, (Space, PH) are uncorrelated.}
    \label{fig:stop_go_correlations}
\end{figure}

The stop \& go regime is now briefly discussed. Fig.\ref{fig:stop_go_correlations}(a) shows the typical dynamics of the swimmer: the swimmer abruptly passes from pure resting states to motion with speed much larger than the threshold value. Orange dots indicate the peaks and the gray areas their widths (delimited by the red dots in the minima adjacent the maxima).
Fig.\ref{fig:stop_go_correlations}(b) reports the average of the lag-1 correlation (that is the correlation between a peak and its first neighbor) for the set of W/G=4/1 composition (the trend is similar for other viscosities). While the calculated mean lag-1 correlation for the displacement ranges between 0.2 and 0.5 (depending of the W/G ratio), the large CV (typically $CV>1$) renders the mean physically unrepresentative. This suggests that the propulsion mechanism, driven by the ME from an homogeneous surfactant source, is dominated by stochastic symmetry-breaking. The lack of a stable correlation confirms that the rest period (typically between 20 and 40 s), combined with the repulsive boundary interactions with the pool's edges, effectively decouples the chemical and mechanical state of the swimmer from its previous peak. This decoupling is further supported by the separation of time scales between the active pulses and fluid relaxation. Given the kinematic viscosity of the W/G=4/1 mixture ($\nu \sim 1.9 \cdot 10^{-6}\ m^{2}/s$), the characteristic momentum diffusion time across the disk radius ($R=5\ mm$) is $t_{diff} = R^{2}/\nu \sim 10\ s$. Since the inter-peak delay significantly exceeds this dissipation threshold, the local hydrodynamic field is allowed to fully relax between events. Consequently, each motion event is reset by the combination of full viscous dissipation and the randomizing effects of local surface tension fluctuations, ensuring that each spike originates from a nearly quiescent hydrodynamic state.

\subsection*{Modeling}
As noted above, the complexity of the system prevents the possibility of performing an accurate and exhaustive modeling that quantitatively considers all the physical effects that concur in the experiments. Although the stop \& go regime has already been discussed in the literature and is caused by stochastically generated asymmetries in the released surfactant wake, the sustained oscillations overlapped with the exponential decay is a completely new phenomenon. In fact they were cited in \cite{PhysRevResearch.2.043299,AKELLA20181176}  but their origin has been never discussed before.\\
The model assumes an extended source with a Gaussian spatial profile and a temporal exponential decay of surfactant flow. Its initial speed creates the concentration asymmetry required to initiate the motion. The source effectively couples the instantaneous Marangoni propulsive force (which is proportional to the surfactant concentration gradient) to the diffusion equation governing the chemical field. This simplification is strongly supported by the experimental evidence: the swimmer initially follows a straight path (only tracking the circular pool edge due to liquid meniscus-induced confinement). This indicates that the flow-induced "noise" (the combined advective and diffusive contributions to the liquid flow) is negligible during the initial phase, confirming that inertial effects overwhelmingly dominate the overall dynamics in this oscillatory regime.\\
Experimentally, the forces that act simultaneously during the swimmer motion are:
\begin{enumerate}
\item the propelling force due to the ME that should decrease with time proportionally to the ethanol flux ($F_{M}$);
\item a resistive term due to the drag generated by the viscosity of the solution ($F_{d}$);
\item the edge effect of the finite dimension of the pool ($F_{curv}$). This is created by the convex shape of the liquid meniscus acting as a repulsive potential well that keeps the swimmer entrapped inside the pool and prevents any direct interaction with its edges (as can be seen in Fig. \ref{fig:speed_vs_time}(b-top)).
\end{enumerate}
These three forces sum up to provide the overall resultant:
\begin{equation}
    F_{net} = m \frac{dv}{dt} = F_{M} + F_{d} + F_{curv}
\end{equation}

Since a simplified model able to capture the basic physics of the system is required, the $F_{curv}$ term has been assumed constant (given the rounded shape of the pool) and is included in a generalized $F_{d}$.\\
The 1D reaction-diffusion self-propulsion model considers two major sources of force:
\begin{enumerate}
    \item the propelling Marangoni effect is described phenomenologically with:
    \begin{equation}
        F_{M} = \phi(x,t) \cdot \frac{dC(x,t)}{dt}
    \end{equation}
    where $\phi(x,t)$ is the flux of ethanol and $C(x,t)$ is the spatial concentration profile of ethanol.
    \item a single drag term accounting for both viscous drag and meniscus-induced confinement.
    The actual dependence of the drag coefficient ($c_{D}$) from $Re$ has a complex shape \cite{Panton2013}. In laminar flow, it is often approximated by $24/Re$, while in the considered regime several empirical corrections were proposed, slightly departing from the linear behavior and require experimental parameters (density and viscosity) which cannot be properly estimate for the 1D model \cite{drag_coeff}. Thus, a linear approximation is assumed considering:
    \begin{equation}
        F_{D} = -\frac{1}{2} \rho v^{2} c_{D} A \propto -\alpha \frac{v^{2}}{Re} \sim -\beta \frac{v^2}{\mathbf{v}} \sim -\gamma \mathbf{v}
    \end{equation}
    where: $\rho$ is the fluid density and $A$ is the characteristic cross section exposed to the fluid.
    The departure of the linear approximation from the empirical models accounts for up to few percent of difference in the drag force (see Fig. 4 of the ESI for a quantitative comparison). Moreover, since the speed of the swimmer in the rare long linear trajectories it traveled in the large pool have speeds comparable to those of the circular ones, it can be deduced that the confinement effect produces a small force and that its dependence on $Re$ can be accounted for in the single phenomenological drag term.
\end{enumerate}
The system of equations is composed of two ODE ($x_{s}$ and $v_{s}$) coupled to the PDE $dC/dx$ term. The PDE equation is coupled back to the ODE system via $x_{s}(x,t)$. The swimmer is defined as an object of finite size, of characteristic width $\sigma_{s}$ (see below), which releases ethanol with an exponentially decaying time profile:
\begin{equation}
    \phi(x_{s},t) = \phi_{0} \cdot exp \left( -\frac{t}{\tau_{rel}} \right)\\
\end{equation}
where $\phi(x_{s},t)$ is the flux at the current swimmer position $x_{s}$, $\phi_{0}$ is the flux at $t=0$ and $\tau_{rel}$ is the decay time of the flux.\\
At $t=0$ the swimmer has a small but finite speed ($\mathbf{v}(x_{0}, t_{0}) = 0.1$) that imposes the asymmetric ethanol concentration profile, $C(x,t)$, required to generate the $F_{M}$ term. Spatially $C(x,t)$ is defined by considering the swimmer as a gaussian source of width $\sigma_{s}$ and by weighting its flux contribution across the 1D domain using:
\begin{equation}
    J(x,t) = \frac{1}{\sigma_{s} \sqrt{2\pi}} \cdot exp \left( - \frac{(x - x_{s}(t))^{2}}{2\sigma_{s}^2}\right)\cdot \phi(x_{s},t)
\end{equation}
$J$ is the contribution of the source to the concentration profile $C(x,t)$. $J$ weights the flux over a smoothly extended region both to consider the finite size effect of the swimmer and to avoid the numerical problems of integrating sharp interfaces or point-like sources \cite{Boniface2019Selfpropulsion}. The differential equation determining $C(x,t)$ is made by three terms:
\begin{equation}
    \frac{dC(x,t)}{dt} = D \cdot \nabla^2 C(x,t) - \frac{C(x,t)}{\tau_d} + J(x,t)
\end{equation}
the first term is the solution to the diffusion equation (assuming constant diffusion coefficient $D$), the second mimics the mixing of the ethanol from the air-liquid interface layer (the source of ME) to the bulk of the solution with a characteristic $\tau_{d}$ time constant. The ratio between $D$ and $\tau_{d}$ determines the type of motion and the appearance of the oscillations. Finally, the third term models the flux of ethanol released by the swimmer.\\
Numerically, the 1D domain is modeled as $1001$ points grid mapping the $352\ mm$ circumference of the experimental pool and $\sigma_{s}= 10\ mm$ is the swimmer physical diameter. Apart from these spatial correspondences, the 1D model does not maintain the physical units of the real system and it is unrealistic to map both the system parameters and the results into physically relevant units. Yet, the model demonstrates the interaction of the swimmer propelling along closed loops, with its self-released chemical wake.

\begin{figure}
    \centering
    \includegraphics[width=1.\linewidth]{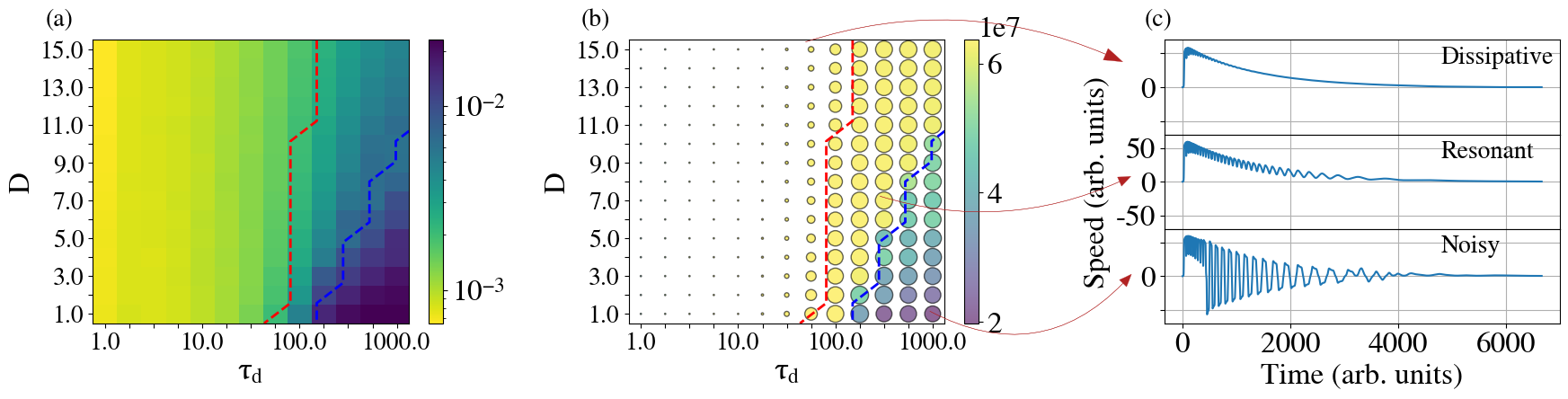}
    \caption{2D maps classifying the different types of motion versus $D$ and $\tau_{d}$: (a) Sample entropy, (b) the colormap indicates the cumulative space traveled by the swimmer during the simulation. The size of the dots is proportional to the number of oscillations: small $\tau_{d}$ produces a strong dampening of the speed oscillations (the dissipative regime to the left of the red dashed line). There is an intermediate and narrow region where sustained oscillations appear for the large part of the dynamics (the resonant regime embedded between red and blue dashed lines). Finally, large $\tau_{d}$ and small $D$ create the conditions for unstable and wandering trajectories (to the right of the dashed blue line). The dashed lines are the boundaries of the motions as classified by the GMM algorithm. (c) Three representative speed profiles acquired from the specific ($D, \tau_{d}$) combinations as indicated by the arrows.}
    \label{fig:classification}
\end{figure}

To classify the regime of motions, three parameters of the motion were considered and are applied to the ($D,\tau_{d}$) 2D map:
\begin{enumerate}
    \item the Sample Entropy (SE) to assess the complexity of time-series signals: higher SE indicates less self-similarity of the template vectors and more noisy trajectories;
    \item the Number of Oscillations (NO), before the breakpoint, to check the length of the resonant state;
    \item the Accumulated Traveled Space (ATS). this is the algebraic cumulative sum of the space traveled by the swimmer: noisy states produce changes of sign of the speed that decreases the overall ATS and, thus, correlates with a noisy state. 
\end{enumerate}

\begin{figure}
    \centering
    \includegraphics[width=0.7\linewidth]{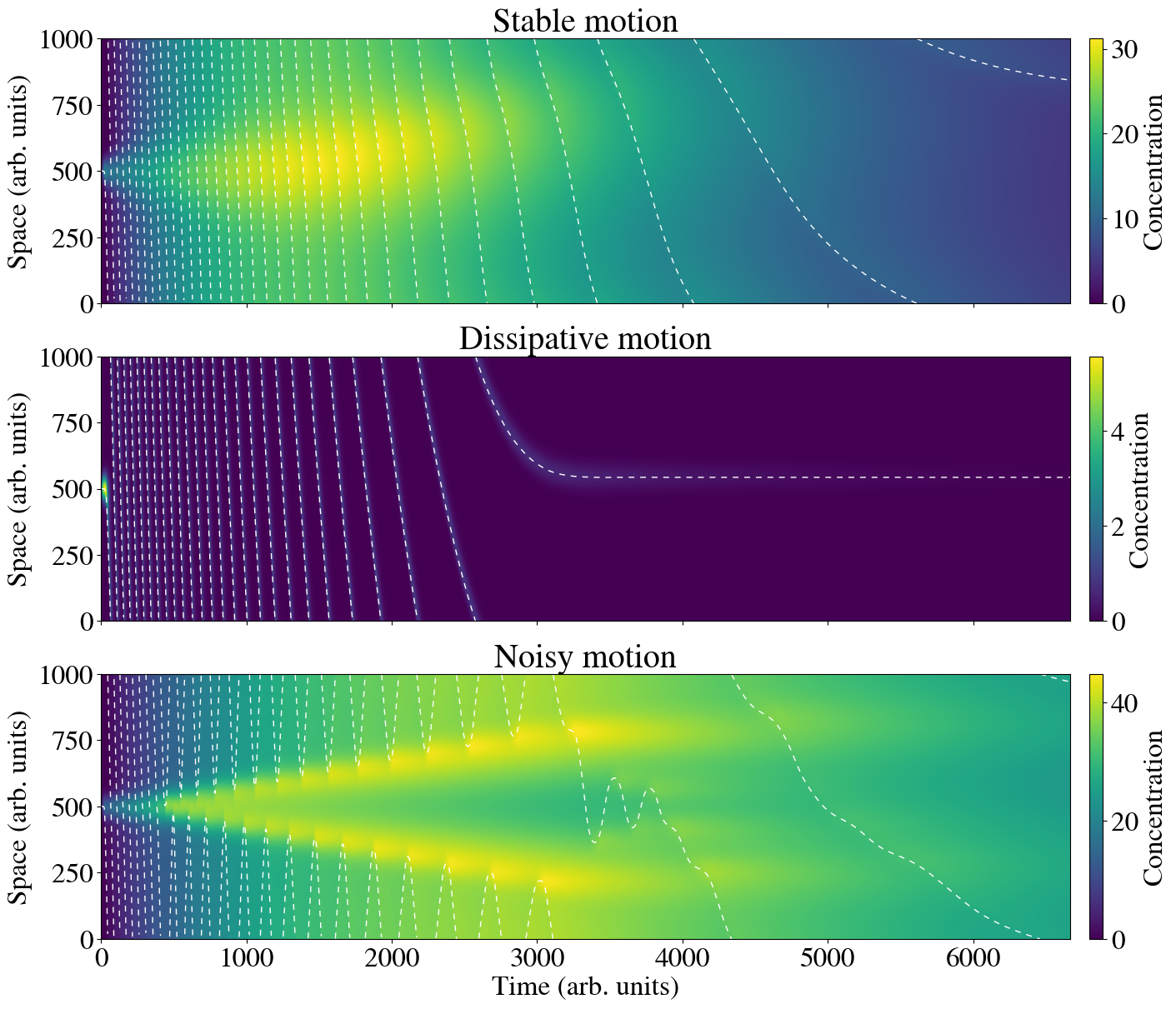}
    \caption{Concentration maps for the three types of motion. The colormap indicates the concentration profile of the surfactant along the 1D simulation domain. The white dotted line indicates the position of the swimmer: the motion starts at the center of the domain and proceeds downward. (top) The stable state: the speed of the swimmer and the dynamic of dissipation of the high concentration cloud enable the formation of a \emph{"resonant"} state where the balance between the inertia of the swimmer and the mixing of the surface released surfactant balance properly. (middle) a larger $D$ strongly reduces the overall concentration gradient. The small residual halo around the swimmer is enough to propel it but does not last for an entire period of propagation around the pool circumference. (bottom) small $D$ and large $\tau_{d}$ create instabilities and multiple maxima of concentration that render the motion chaotic.}
    \label{fig:C_maps_types_motion}
\end{figure}

The three types of motion are robustly recognized using the unsupervised Gaussian mixture model (GMM) classification algorithm and the boundaries are illustrated by the red and blue lines in Fig. \ref{fig:classification}(a,b). After normalizing and rescaling the features (SE, NO, ATS), ATS were converted into a binary mask so that each $\vec{v}(x,t)$ trace containing negative speed larger than 10\% of its $Max(|\vec{v}(x,t)|)$ is considered as False. Such empirical transformation allows the GMM to correctly classify the motion types without tuning the weight of any space feature parameters. Fig.\ref{fig:classification}(a) shows the SE map: small $\tau_{d}$ (that is long mixing time between the surface layer and the bulk of the solution) always create a dissipative regime with smooth speed profiles having no or highly damped oscillations, as shown in Fig.\ref{fig:classification}(c-top). As $\tau_{d}$ increases, there is a narrow regime where speed oscillations are sustained for a substantial part of the motion (Fig.\ref{fig:classification}(c-middle)). Similarly, Fig.\ref{fig:classification}(b) indicates the other two parameters used to classify the regions of motion: the colorbar shows the ATS traveled by the swimmer. While smooth trajectories maintain the sign of the speed and increase their ATS, noisy states have smaller ATS due to speed inversions. The size of the dot is proportional to the NO measured along the $\vec{v}(t)$ vector. Given a fixed $D$, small $\tau_{d}$ allows the concentration of ethanol to accumulate around the swimmer, at a fast rate and to saturate without producing the spatially localized cloud that is responsible for the oscillations. However, a large $\tau_{d}$ produces strong asymmetries in the concentration profile that make the motion noisy (Fig.\ref{fig:classification}(c-bottom)). In the intermediate regime, the feedback provided by the cloud to the swimmer sustains chirped oscillations that might last for the entire duration of the smooth regime of motion.\\
The three categories of motion are clearly visualized by the concentration maps superimposed onto the position of the swimmer as in Fig.\ref{fig:C_maps_types_motion}.
Fig.\ref{fig:C_maps_types_motion}(top) shows how the cloud of surfactant released by the swimmer evolves in time and space in a \emph{resonant} state ($D=7.0 \text{ and } \tau_{d}=316.23$). The white dashed line indicates the position of the swimmer as it cycles along the 1D domain. Note that the initial speed of the swimmer (fixed at $\vec{v} = + 0.1 \hat{x} $) always produces an inversion of its speed so that it propels towards negative direction. The slowing down of the swimmer during the experiment produces two effects: an obvious decrease of the steepness of the white traces as time goes by and an apparent motion towards positive coordinates of the ethanol concentration cloud. This regime satisfies a delicate balance so that: 1) the inertia of the swimmer is large enough to cross the high concentration region (which tends to slow it down) and 2) the spatial gradient of the surfactant cloud is large enough to provide the required momentum to cross the cloud at the next loop. Fig.\ref{fig:C_maps_types_motion}(central) shows the evolution of a \emph{dissipative} state ($D=14.0 \text{ and } \tau_{d}=1.78$). In this case, by increasing the ratio $D/\tau_{d}$, the overall concentration gradient decreases and localizes around the swimmer position. The absence of a persistent cloud of large concentration prevents the generation of the feedback mechanism that sustains the speed oscillations. Moreover, the fast diffusion of the surfactant is effective in mixing and homogenizing its concentration throughout the simulation domain, so that the swimmer trajectory overlaps with a faint increase in the surfactant concentration that localizes around the swimmer and the motion runs out early. Finally, Fig.\ref{fig:C_maps_types_motion}(bottom) shows the evolution of a \emph{noisy} state ($D=1.0 \text{ and } \tau_{d}=1000.0$). In this case, the swimmer is initially propelled by the ME but its large inertia push it beyond the trailing edge of the cloud during the very first loops, thus allowing the swimmer to escape from the cloud. Given the small value of $D$, the cloud slowly diffuses and the large $\tau_{d}$ works as an effective sink to remove the already released surfactant. However, a concentration gradient still piles up behind the swimmer by the surfactant it releases in the last instants. When the concentration gradient is large enough, it works as an effective barrier that repels the swimmer and invert its propagation direction. The cloud tends to disappear in a short time but its persistence is long enough to slow down the incoming swimmer. This fact increases the concentration gradient in the reinforced cloud and continuously repels the swimmer. Such a dynamic creates a metastable state with a nearly symmetric configuration of two high concentration regions acting as repulsive barriers for the swimmer. Subsequently, as the inertial effects decrease, small perturbations disrupt the initial symmetric configuration, causing the trajectory to transition into chaotic dynamics.

\section*{Conclusions}
This work characterized the dynamics of macroscopic hydrogel swimmers in the intermediate Reynolds regime ($Re \sim 100-300$), uncovering a fundamental transition from resonant oscillatory motion to a stochastic stop \& go behavior. The results establish that the sustained speed oscillations observed during the initial exponential decay phase are the product of a dynamic self-interaction between the swimmer and its own persistent chemical wake. This chemical self-feedback, successfully reproduced using a simplified 1D reaction-diffusion model, is governed by the delicate balance between chemical diffusion and mixing scales ($D$ and $\tau_{d}$). Crucially, a universal threshold speed of approximately $10\pm 2.5\ mm/s$ is identified, which remains consistent across different solvent viscosities, marking the point where inertial self-caging fails and the system transitions into a chaotic, decoupled state. By elucidating how hydrodynamic inertia, chemical fields, and spatial confinement couple to determine motion properties, this study provides predictive tools and a theoretical foundation for the design of stable, high-speed active matter systems in macro-scale applications.

\section*{Methods}
\subsection*{Synthesis of Nanocellulose}
The procedure exploited to obtain CNC is a slightly modified version of \cite{Maestri2017Fabrication}. The starting material for the production of cellulose nanocrystals (CNC) is never-dried soft bleached pulp (Celeste85 from SCA \"Ostrand, Sweden). 10 g of cellulose pulp are added in a beaker to 500 mL of water and sonicated. 1 g of NaBr and 160 mg of (2,2,6,6-Tetramethylpiperidin-1-yl)oxyl (TEMPO) are dissolved in 100 mL - 150 mL of water. The two solutions are merged, and the total volume is brought to 1 L. The suspension is added with 35 mL of NaClO (nominal 6-14\% active chlorine), the pH is then maintained to about 10-11 by subsequent additions of 1 mL of NaOH 1M. The reaction lasts for a few hours until the pH remains constant, indicating total oxidation of the anhydroglucose units $C_{6}$ site. The suspension is left to rest, the clear supernatant is removed and then restored with deionized water. The process is repeated until the pH reaches a neutral value. Typically, the process produces $\sim 800\ mL$ of concentrated CNC suspension.\\
35 mL of the concentrated CNC suspension are collected under vigorous stirring and poured into a plastic tube, which is sonicated at 20 kHz and 100 W output power twice for 2 min (waiting for about 10 min between each sonication, to avoid excessive heating), while kept in an ice bath. The final slurry is then filtered to remove any impurities (such as the titanium residues released by the tip of the homogenizer).\\
The CNC concentration is measured by letting a known volume V (3 mL) to evaporate in an oven at about $60^{\circ} C$ overnight. The residual mass allows for the estimation of the CNC concentration that is, generally, in the range 3-4 mg/mL right after the synthesis. The suspension are concentrated using a rotary evaporator until it reaches concentration of 7-9 mg/mL.
\subsection*{Ionotropic gelation}
Generally, ionotropic gelation is achieved either by dropping an ionic solution into a polymer precursor (direct gelation) or by reversing these roles in the case of inverse gelation. Both methods allow for good control over the size and shape of the gels, but do not ensure good homogeneity of bulk materials. In fact, given the nearly instantaneous gelation, the region where the drop impacts the solution \emph{freeze} the CNC suspension into volumes with strong non-equilibrium concentration of cations, their diffusion is hampered and might not bring to equilibrium structure in reasonable times. To avoid this issue, a custom device was developed that takes advantage of the diffusion of cations from a highly concentrated aqueous solution of CaCl$_{2}$ into the CNC suspension, via a dialysis membrane. This fact ensures the formation of a more homogeneous gel structure since the cations slowly diffuse into the CNC suspension, giving time to the gel to equilibrate its structure during the gelation. The device is sketched in Fig. 5 of the ESI.
The gelation reaction is left to proceed overnight and each batch of swimmers is used the same day, avoiding their storage to reduce interferences from possible long term gel dynamics. Since a colored swimmer is needed for its subsequent optical tracking, after gelation they are loaded for two hours in ethanol solution colored with methyl blue (from Sigma-Aldrich).

\subsection*{Data Analysis and visualization}
\subsubsection*{Tracking of the swimmer and analysis of the trajectories}
Swimmer motion was acquired using a digital camera (Panasonic Mod. DC-TZ95 res. 1920x1080 @ 30fps). The camera is placed on top of a Teflon pool of 118 mm of internal diameter. The height of the liquid was fixed at 16 mm, the value is chosen so that the liquid meniscus assumes a convex shape that effectively confines the swimmer within the pool and avoids the direct interaction with the solid wall. To investigate how the viscosity of the liquid affects the motion of the swimmer the pool was filled with three different water/glycerol (W/G) solutions with volume ratio of 3/1, 4/1 and 5/1.\\
Post-processing was conducted using a custom Python pipeline: a python script detects the position of both the pool and of the swimmer and returns their cartesian coordinates for the entire duration of the experiment (which lasts for about 30 min). Another script analyzes the trajectory and creates the paper figures. All the analysis were performed using: NumPy\cite{numpy}, SciPy\cite{scipy}, Scikit-Learn\cite{sklearn} and PyWavelets\cite{pywavelets}. All figures were generated using Matplotlib\cite{matplotlib}.

\subsubsection*{Modeling of the swimmer}
The routine to perform the modeling has been developed in Julia\cite{julia} programming language. This choice was dictated by the need for a fast programming language that allows to perform several thousands of simulation to span across the $D,\ \tau_{d}$ space and find the optimal conditions for oscillating behavior. The Julia module implements a high-fidelity 1D simulation of a Marangoni swimmer by coupling the mechanical dynamics of an object with an evolving chemical concentration field through a system of ODEs. The core physics involves a self-propelled disk that releases a surfactant, generating surface tension gradients-calculated using Automatic Differentiation via ForwardDiff.jl\cite{automaticdiff} to produce a Marangoni force that competes with viscous drag. The concentration profile evolves on a periodic grid subject to diffusion, Gaussian sourcing, and linear decay, with numerical integration handled by the DifferentialEquations.jl\cite{differentialequations} suite. The integration scheme is AutoVern9(Rodas5()), which dynamically switches between explicit and implicit approaches depending on the local behavior of the swimmer and the concentration field. In fact, while the swimmer moves at a slow speed, the chemical gradients might change almost instantly, defining the stiffness of the problem.

\bibliography{sample}

\section*{Acknowledgements}
The authors acknowledge prof. M. Scarpa and Dr. E. D'Amato for their valuable support and SCA Ostrand (Sweden) for the supply of cellulose raw material.

\section*{Author contributions statement}
A.F. Methodology, Software, Formal analysis, Investigation, Data Curation, Writing - Original Draft. P.B. Conceptualization, Methodology, Software, Validation, Formal analysis, Investigation, Resources, Data Curation, Writing - Original Draft, Visualization, Supervision, Project administration, Funding acquisition

\section*{Competing Interests}
The authors declare no competing interests.

\section*{Data Availability}
The datasets generated during during the current study are available in the [NAME] repository, [PERSISTENT WEB LINK TO DATASETS]. The python code used to analyze the data and the julia code use to run the modeling are available in the [NAME] repository, [PERSISTENT WEB LINK TO DATASETS].

\newpage

\appendix
\begin{center}
    
    {\huge \textbf{Inertial Self-Caging: Dynamics of Macroscopic Swimmers at moderate Reynolds Number Sustaining Chemical Wake Resonance}} \\
    \vspace{1.5cm}
    
    {\Large Alessandro Foradori$^{1}$ and Paolo Bettotti$^{1,*}$} \\
    \vspace{0.8cm}
    
    {\small $^{1}$Department of Physics, Nanoscience Laboratory, University of Trento, \\ via Sommarive 14, 38123 Povo (TN), Italy} \\
    \vspace{1.5cm}
    
    \rule{0.5\textwidth}{0.4pt} \\
    \vspace{0.3cm}
    {\small *Email: \href{mailto:paolo.bettotti@unitn.it}{paolo.bettotti@unitn.it}}
\end{center}

\vspace{3cm}

\vfill

\newpage

\section{Detection of the Breakpoint}
\begin{figure}[H]
    \centering
    \includegraphics[width=0.75\linewidth]{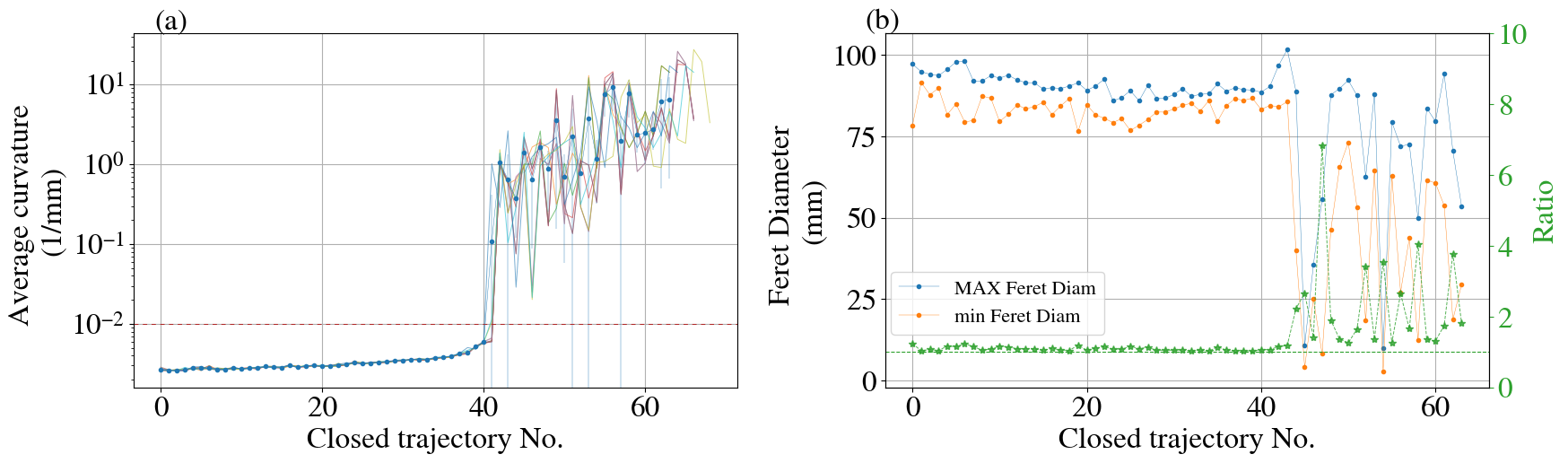}
    \caption{Characteristic parameters to distinguish between the initial region of smooth inertial motion and the stop \& go one. (a) Colored lines show the average curvature of each closed trajectory, while the blue dotted line is the average over the shifted definition of closed loops. (b) Maximum (blue dotted line) and minimum (orange dotted line) Feret diameters and their ratio (green).}
    \label{fig:closed_traj}
\end{figure}
The following strategies are both effective and robust in determining the breakpoint:
\begin{enumerate}
    \item the local curvature of the trajectory is nearly constant in the oscillating regime and changes to a highly disordered one in the last part of the experiments. Curvature is defined as:
    \begin{equation}
        \kappa = \frac{|x'y'' - y'x''|}{(x'^{2}+y'^{2})^{3/2}}
    \end{equation}
    where the prime functions indicated the derivative of the swimmer displacement along the specific cartesian axis;
    \item the ratio of the Feret diameters, which are defined as the extreme lengths describing the convex hull of each closed loop. This ratio remains nearly constant (approximately $\sim 1$) in the first part of the experiments, indicating highly symmetric loops, but becomes significantly noisier toward the end.
\end{enumerate}
The two approaches defined above have a significant advantage compared to the standard ones. In fact, a closed loop trajectory is defined by measuring the movement vector between adjacent videoframes with respect to a reference axis (eg. $\hat{u_{x}}$: every time this vector wrap $2\pi$ a closed loop is swept). Since the first step can be arbitrarily chosen along any part of the smooth circular path, and that the speed of the swimmer does not decrease by a large amount during the entire experiment, we consider the number of steps required to traverse the first loop (which is the shortest in time) and divide it into $n$ equally spaced segments (typically $n = 10$). By repeating the analysis of both the curvature and the ratio of the Feret diameters and by shifting the first point by the amount of steps defined by $n$, we effectively averaging the values across local sections of the trajectories starting at slightly shifted times and space position.\\
Fig. \ref{fig:closed_traj}(a) reports the average curvature of each closed trajectory: the different colored lines show the curvature for closed loops defined by shifting their initial point as described above. The blue dotted line is the average over the $n$ definitions of closed loops. Similarly, the ratio of the Feret diameters can also be used to define the change point separating the two regimes of motion. Since the variation in the average curvature is larger than the variation of the ratio of the Feret diameters, it has been selected as candidate parameter to classify the threshold time. Empirically, we found that a value of the curvature of $10^{-2}\ mm^{-1}$ is an effective threshold for all fluid compositions investigated (different water/glycerol ratio) and has been chosen to define the time at which the motion change: the first trajectory having local curvature above such threshold defines the start of the second part of the motion, dominated by chaotic accelerations.

\newpage
\section{Typical dynamics at larger viscosities}
\begin{figure}
    \centering
    \includegraphics[width=1\columnwidth]{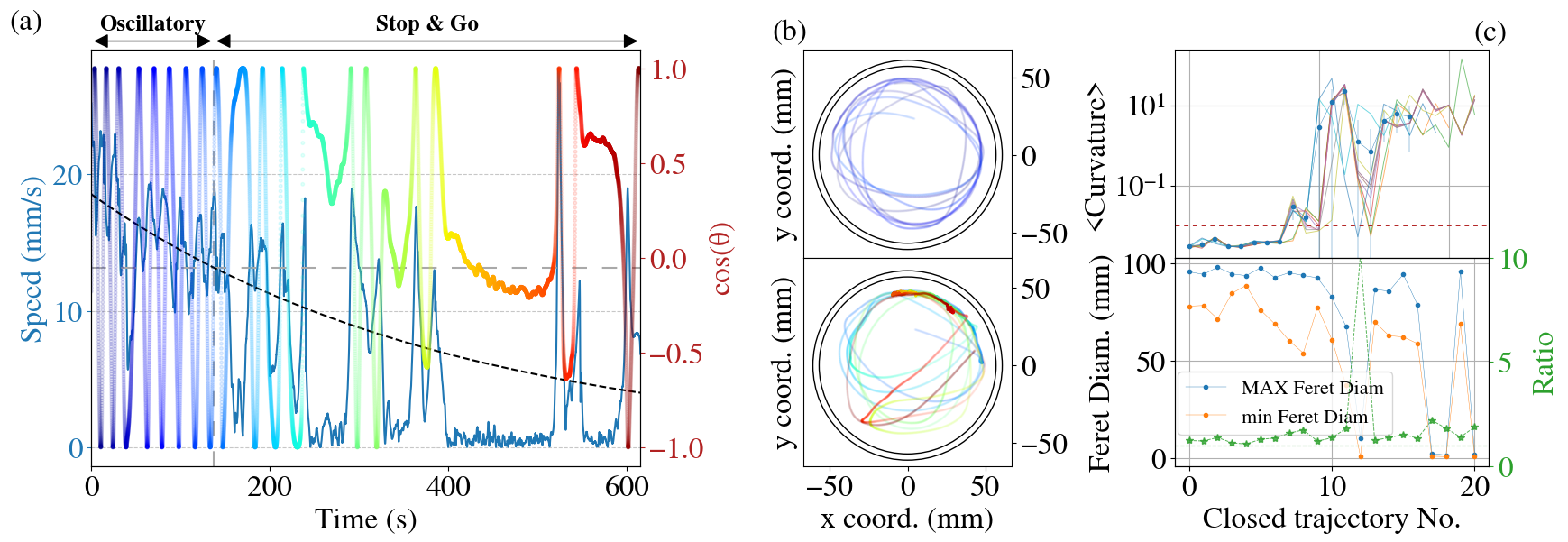}\newline
    \includegraphics[width=1\columnwidth]{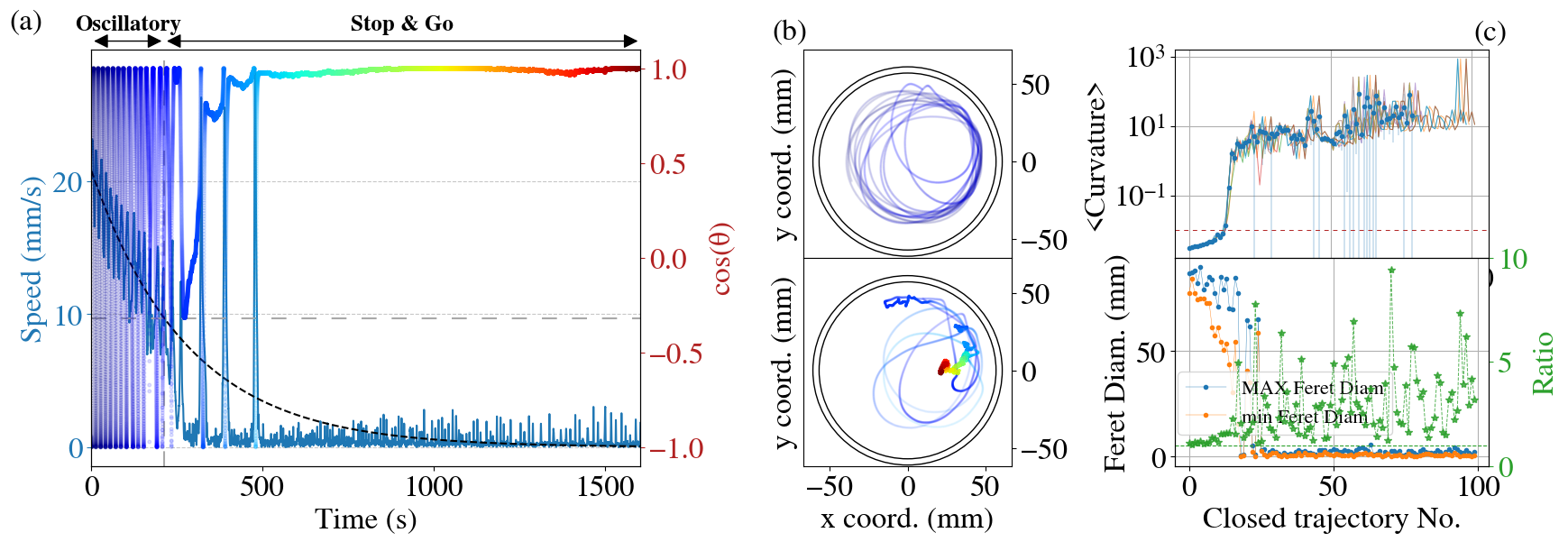}\newline
    \includegraphics[width=1\columnwidth]{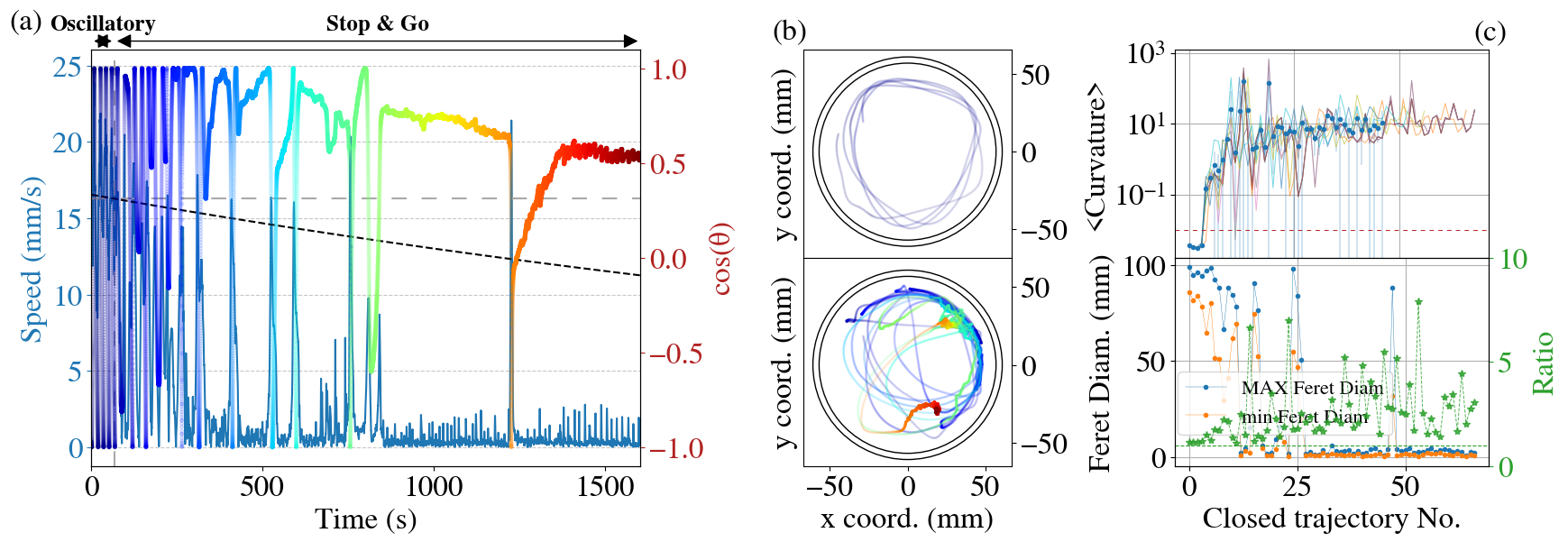}\newline
    \caption{Examples of trajectories for swimmers moving in a W/G=1/1 ratio. It is clear that the oscillating regime happens only during the first instances of the motion and most of the time the swimmer wonders following stochastics trajectories.}
    \label{fig:ratio1}
\end{figure}

\begin{figure}
    \centering
    \includegraphics[width=1\columnwidth]{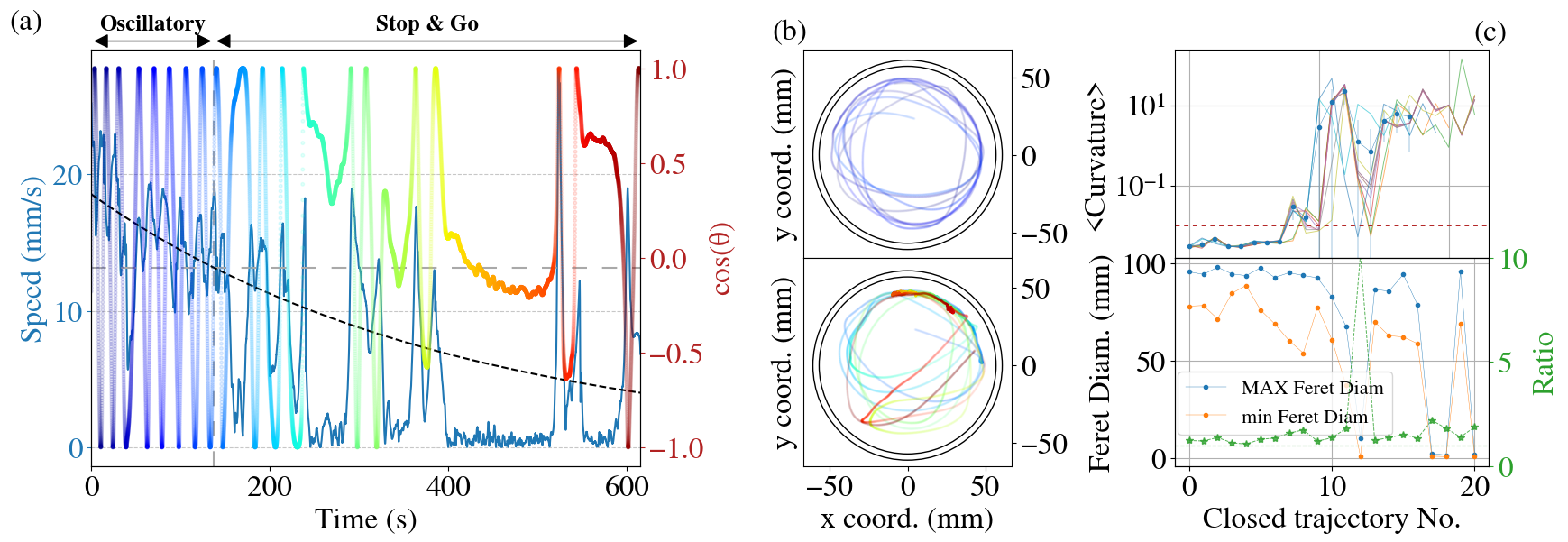}\newline
    \includegraphics[width=1\columnwidth]{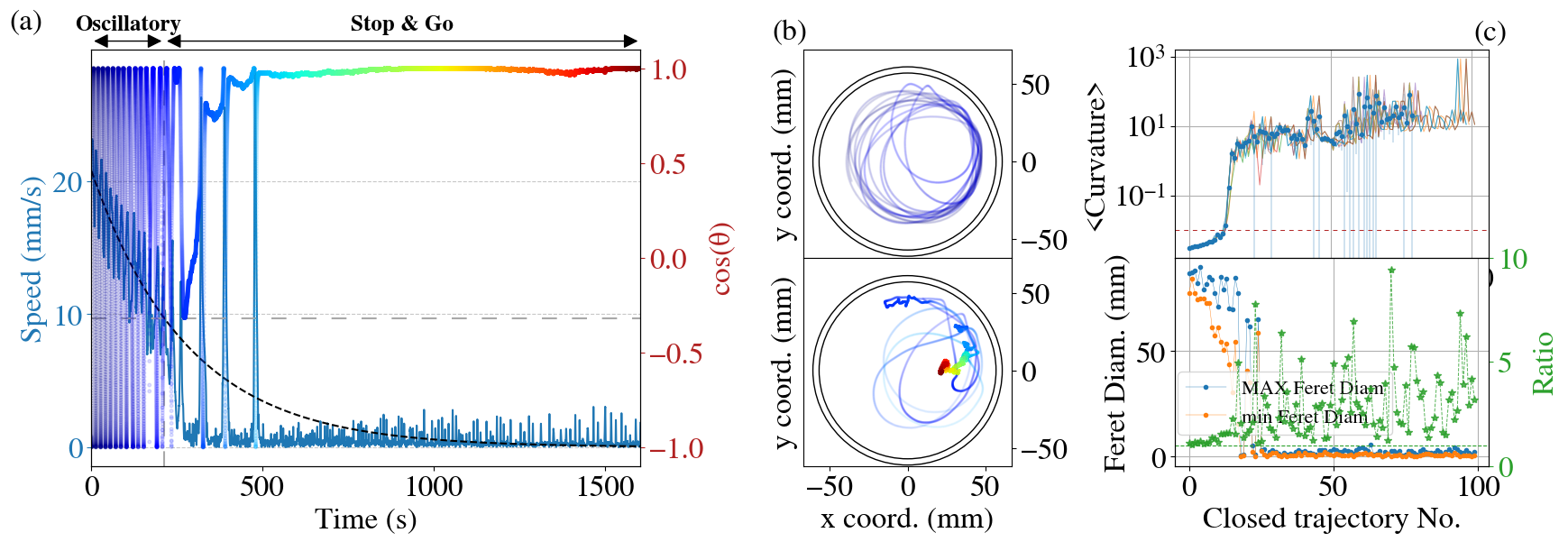}\newline
    \includegraphics[width=1\columnwidth]{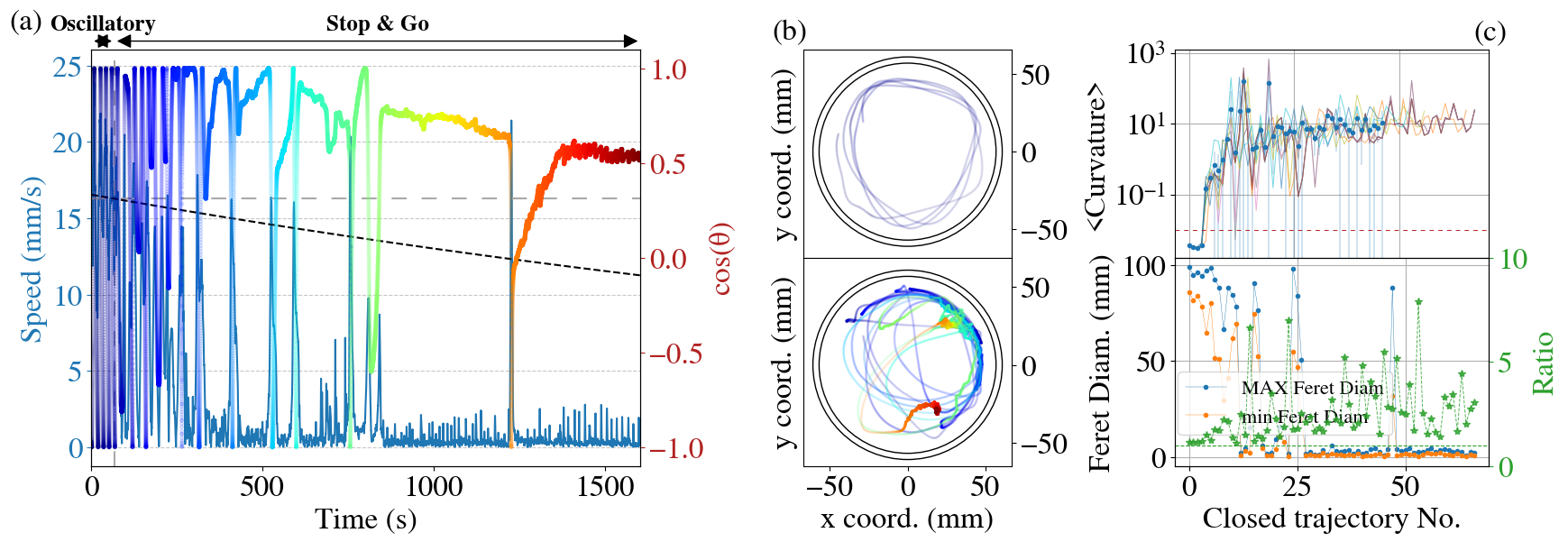}\newline
    \caption{Examples of trajectories for swimmers moving in a W/G=1.74/1 ratio. This ratio is the one that doubles the viscosity of the W/G=1/1 case. As in the previous case, in most of the experiments the motion cannot enter into the resonant state.}
    \label{fig:ratio1.74}
\end{figure}

Figs. \ref{fig:ratio1}-\ref{fig:ratio1.74} show typical trajectories for swimmers released in solutions of grater viscosity W/G=1/1 and W/G=1/1.74 (the viscosity of the latter is twice the one of the former): in both cases the swimmer performs only a few circular loops (if any) of reduced circularity, as can be seen by looking at the significant differences between the Feret diameters. Moreover, the stop \& go regime features only sparse peaks. This is an expected behavior caused by the following two reasons:
\begin{enumerate}
\item the swimmer does not have enough energy to sustain the oscillating motion for prolonged times. This fact arise from the nonlinear behavior of the viscosity of the water/glycerol system. Table \ref{tab:viscosities} resumes how concentrations and viscosity change. While the mass increase of glycerin is about of the 25\%, the kinematic viscosity triples: the drag force increases dramatically and the \emph{power budget} required to maintain continuous symmetry-breaking becomes higher, too. Since the flux of surfactant decreases exponentially, it ends early below the value required to sustain the motion of the swimmer.
\begin{table}[H]
        \centering
        \begin{tabular}{cccc}
            W/G (v/v) & Vol. G (\%) & Mass G (\%) & $\eta\ (m^{2}/s)$\\
            3/1 & 25 & 30 & $2.3\cdot10^{-6}$\\
            1/1 & 50 & 55 & $7.5\cdot10^{-6}$\\
        \end{tabular}
        \caption{Variation of the kinematic viscosity vs W/G ratio.}
        \label{tab:viscosities}
    \end{table}
\item The swimmer stops moving continuously when the background concentration of surfactant in the pool becomes high enough to reduce the surface tension gradient below a critical threshold. In the more viscous mixtures two effects accelerate this decrease: 1) a reduced diffusivity of the surfactant and 2) a damped dynamic of the wakes generated by the swimmer's motion thus, the surfactant localizes for longer time around the swimmer, saturating the local surface tension.
\end{enumerate}

The net effect is that the diffusion of the surfactant far from the swimmer is strongly reduced and its advective clearance  becomes slower than the chemical accumulation, in these conditions the Peclet number cannot reach the threshold limit to start a new speed peak (similar to what demonstrated in \cite{Michelin2013} for the laminar case). 

\newpage
\section{Check of the validity of the linear approximation for the drag term}
\begin{figure}[H]
    \centering
    \includegraphics[width=0.6\linewidth]{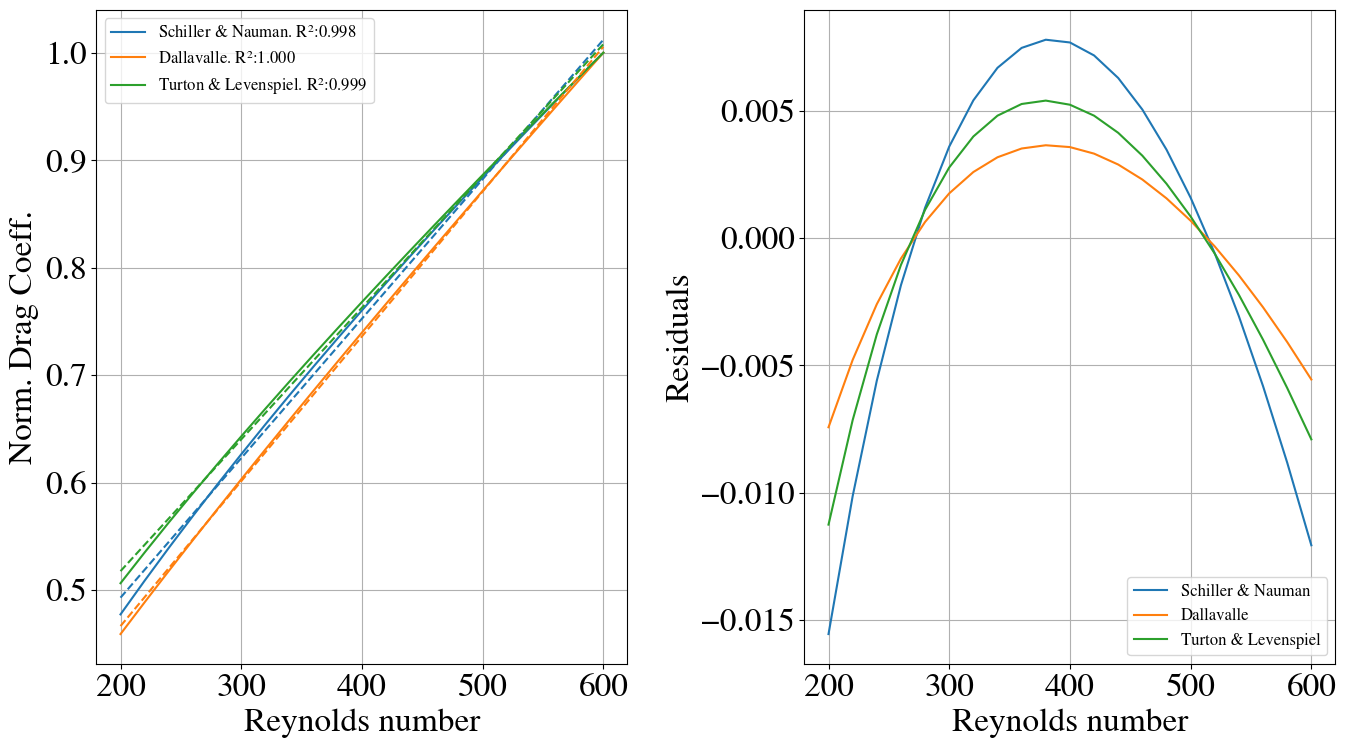}
    \caption{Comparison between empirical models of the drag coefficient ($c_{D}$) in the intermediate Reynolds regime ($Re < 10^{3}$). (a) Empirical equations of normalized $c_{D}$ are fitted with a linear model (dotted lines). (b) residuals between the normalized empirical model and the linear approximation. Models taken from: R. Beetstra, M. A. van der Hoef, and J. A. M. Kuipers, "Drag Force of Intermediate Reynolds Number Flow
    Past Mono- and Bidisperse Arrays of Spheres", AIChE J., 53(2), 489 (2007). doi:10.1002/aic.11065}
    \label{fig:drag_coeff_models}
\end{figure}

Fig.\ref{fig:drag_coeff_models} shows the comparison between well known empirical model to estimate the drag coefficient at intermediate Reynolds number and the linear approximation used in this work.

\newpage
\section{Crafting Tool}
\begin{figure}[H]
    \centering
    \includegraphics[width=0.5\linewidth]{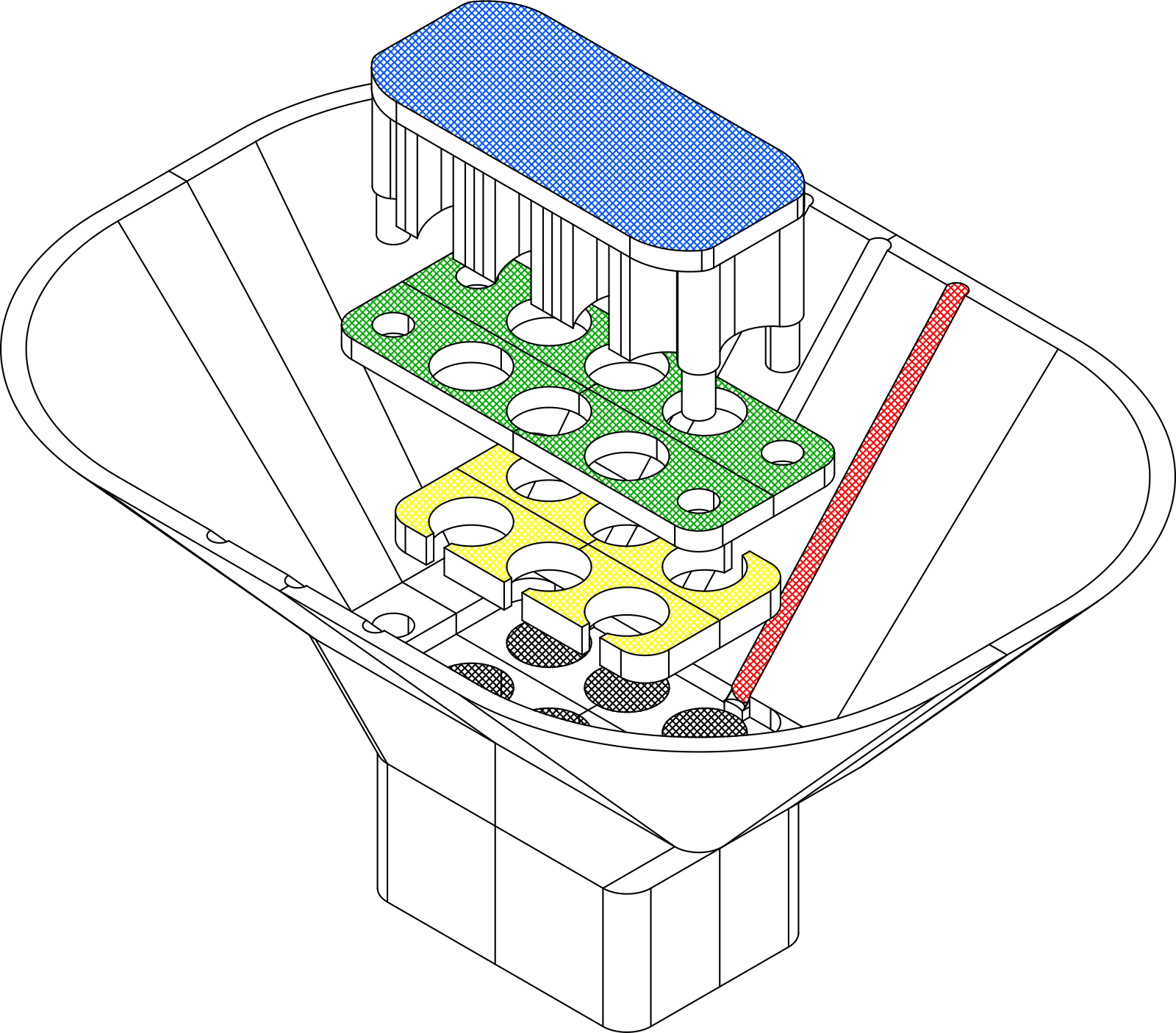}
    \caption{Exploded view of custom tool developed to sculpt the hydrogel disks: the holes on the bottom of the tool (dark spots) connect the $CaCl_{2}$ solution with the CNC suspension. A soft rubber gasket (yellow), with holes of shape similar to that of the hydrogels disks, support the mask (green) that determine the actual shape of the hydrogels (disks in our case). The dialysis membrane is placed in between these two elements. A weight (blue) is placed on top of the stack to assure an hermetic seal. The red channels are fundamental to allow air to escape from the rubber seal and avoid the formation of air bubble right below the dialysis membrane and, thus, its non homogeneous wetting (the figure is taken from: A. Foradori, "Marangoni Propulsion of Nanocellulose Gel", University of Trento, 2023).}
    \label{fig:craftingtool}
\end{figure}

Fig. \ref{fig:craftingtool} shows an exploded 3D view of the custom tool we developed to sculpt the hydrogels with the desired shape. 

\end{document}